\documentclass[fleqn,usenatbib]{mnras}

\usepackage{newtxtext,newtxmath}

\usepackage[T1]{fontenc}
\usepackage{ae,aecompl}

\usepackage{graphicx}	
\usepackage{amsmath}	
\usepackage{amssymb}	
\usepackage{siunitx}    
\usepackage{xcolor}

\newcommand{\Hb}{\rm H\beta}
\defcitealias{Congiu17}{C17}

\title[Radio Structure of Mrk 783]{The Radio Structure of the Narrow-Line Seyfert 1 Mrk 783 with VLBA and e-MERLIN}

\author[E. Congiu et al.]{E. Congiu$^{1,2}$\thanks{E-mail: econgiu@das.uchile.cl},
P. Kharb$^{3}$,
A. Tarchi$^{4}$,
M. Berton$^{5, 6}$,
A. Caccianiga$^{7}$,
S. Chen$^{8}$,
\newauthor
L. Crepaldi$^{9}$,
F. Di Mille$^{2}$,
E. J\"arvel\"a$^{10}$,
M. E. Jarvis$^{11, 12, 13}$,
G. La Mura$^{14}$,
A. Vietri$^{9}$
\\
$^{1}$Departamento de Astronom\'{i}a, Universidad de Chile, Camino del Observatorio 1515, Las Condes, Santiago, Chile;\\
$^{2}$Las Campanas Observatory - Carnegie Institution for Science, Colina el Pino, Casilla 601, La Serena, Chile;\\
$^{3}$National Centre for Radio Astrophysics - Tata Institute of Fundamental Research, Post Bag 3, Ganeshkhind, Pune 411007, India;\\
$^{4}$INAF - Osservatorio Astronomico di Cagliari, Via della Scienza 5, I-09047 Selargius (CA), Italy;\\
$^{5}$Finnish Centre for Astronomy with ESO (FINCA) $-$ University of Turku, Vesilinnantie 5, FIN-20500 University of Turku, Finland;\\
$^{6}$Aalto University Mets\"ahovi Radio Observatory, Mets\"ahovintie 114, FIN-02540 Kylm\"al\"a, Finland;\\
$^{7}$INAF - Osservatorio Astronomico di Brera, Via Brera 28, 20121 Milano, Italy;\\
$^{8}$Physics Department, Technion, Haifa 32000, Israel;\\
$^{9}$Dipartimento di Fisica e Astronomia ``G. Galilei'', Universit\`a di Padova, Vicolo Osservatorio 3, Padova, Italy;\\
$^{10}$European Space Agency, European Space Astronomy Centre, C/Bajo el Castillo s/n, 28692 Villanueva de la Cañada, Madrid,Spain;\\
$^{11}$Max-Planck Institut f\"ur Astrophysik, Karl-Schwarzschild-Str. 1, 85748 Garching, Germany;\\  
$^{12}$European Southern Observatory, Karl-Schwarzschild-str. 2, 85748 Garching, Germany;\\
$^{13}$Ludwig Maximilian Universit\"at, Professor-Huber-Platz 2, 80539 Munich, Germany;\\
$^{14}$LIP - Laboratory of Instrumentation and Experimental Particle Physics, Av. Prof. Gama Pinto 2, 1649-003 Lisboa, Portugal.\\
}
\date{Accepted XXX. Received YYY; in original form ZZZ}

\pubyear{2020}

\begin{document}
\label{firstpage}
\pagerange{\pageref{firstpage}--\pageref{lastpage}}
\maketitle

\begin{abstract}
In this paper, we present the analysis of new radio and optical observations of the narrow-line Seyfert 1 galaxy Mrk 783.
$1.6\,\si{GHz}$ observations performed with the e-MERLIN interferometer confirm the presence of the diffuse emission previously observed.
The Very Long Baseline Array (VLBA) also detects the nuclear source both at $1.6\,\si{GHz}$ (L-Band)  and $5\,\si{GHz}$ (C-band).
While the L-band image shows only an unresolved core, the C-band image shows the presence of a partially resolved structure, at a position angle of $\ang{60}$.
The brightness temperature of the emission in both bands ($>10^6\,\si{K}$) suggests that it is a pc-scale jet produced by the AGN.
The relatively steep VLBA spectral index ($\alpha_{VLBA} = 0.63\pm0.03$) is consistent with the presence of optically thin emission on milliarcsecond scales.
Finally, we investigated two possible scenarios that can result in the misalignment between the kpc and pc-scale radio structure detected in the galaxy.
We also analysed the optical morphology of the galaxy, which suggests that Mrk 783 underwent a merging in relatively recent times.
\end{abstract}




\section{Introduction}
\label{sec:intro}

Narrow-line Seyfert 1 galaxies (NLS1s) are a peculiar class of active galactic nuclei (AGN), characterised by narrow permitted emission lines (full-width at half maximum of $\Hb < 2000\,\si{km.s^{-1}}$), weak Oxygen forbidden lines ([\ion{O}{III}]$\lambda5007/\Hb < 3$) \citep{Osterbrock85} and usually strong Fe II lines \citep{Goodrich89}.
They are believed to be powered by supermassive black holes (SMBHs) with masses around $10^5$ -- $10^8\,\si{M_\odot}$, typically smaller than what is found in normal Seyfert galaxies ($10^7$ -- $10^9\,\si{M_\odot}$) and blazars ($10^8$ -- $10^{10}\,\si{M_\odot}$) \citep{Grupe00,Jarvela15,Cracco16,Chen18}, and they show high Eddington ratios \citep{Boroson92,Sulentic00}.
For these reasons they are often considered to be newly born AGN \citep{Mathur00, Berton17}.

In this paper, we report new enhanced Multi Element Remotely Linked Interferometer Network (e-MERLIN) and Very Long Baseline Array (VLBA) observations of the NLS1 Mrk 783.
The galaxy was observed in a Karl G. Jansky Very Large Array (VLA) radio survey of NLS1 galaxies described in \citet{Berton18}.
During the preliminary phases of the data analysis \citet{Congiu17} (from now on \citetalias{Congiu17}) reported the discovery of a kiloparsec-scale radio (KSR) structure with very steep in-band spectral indexes ($\alpha_{VLA} = 2.02$ at $5\,\si{GHz}$).
The KSR structure was tentatively classified as relic emission.
Mrk 783 is one of the few NLS1s with a KSR structure and the first one where this emission has been tentatively classified as a relic.
Relics are what remains of past activity episodes of an AGN which was able to produce jets.
When the AGN (and/or the jet) turns off, the plasma becomes a reservoir of energy and it can continue emitting at radio wavelength for a relatively long time after the end of the activity phase.
In some cases, relics from multiple activity periods can be observed at the same time in a single galaxy \citep[an excellent case for this is Mrk 6,][]{Kharb06}.
It is now mostly accepted that the AGN phase is only a short period of time in the life of a galaxy, and that it can repeat itself over time in what is known as the duty cycle \citep{Sanders84,Haehnelt93,Czerny09,Sebastian20}.
In this scenario, relics are a fundamental tool in studying the life of AGN and their host galaxies because they give us a view of what has happened in the past and they allow us to investigate how the duty cycle works, its timescales and its impact on the host galaxy \citep{Parma99,Kharb16,Turner18,Brienza18,Sebastian19}.
This could be particularly important for NLS1s.
Relics could help us unveil how these objects, often considered young AGN, are born and how they evolve.

The paper is structured as follows.
In Sec.\,\ref{sec:mrk783}, we shortly describe the main properties of Mrk 783.
Sec.\,\ref{sec:data} describes the reduction of the new radio observations while Sec.\,\ref{sec:results} contains the main results of the data analysis, which are then discussed in Sec.\,\ref{sec:discussion}.
Finally, in Sec.\,\ref{sec:conclusions} we provide a summary of the work.
In this paper we adopt a standard $\Lambda$CDM cosmology, with $H_0 = 70\,\si{km.s^{-1}.Mpc^{-1}}$ and $\Omega_{\Lambda} = 0.73$ \citep{Komatsu11} and spectral indexes $\alpha$ are defined as $S_{\nu} \propto \nu^{-\alpha}$, where $S_{\nu}$ is the flux density of the object at the frequency $\nu$.

\section{Mrk 783}
\label{sec:mrk783}

Mrk\,783 \citep[R.A. = $13$h $02$m $58.8$s Dec=$+16$d $24$m $27$s, z = 0.0672;][]{Hewitt91} was first classified as a NLS1 by \citet{Osterbrock85}.
The mass of the central SMBH is $4\times10^7\,\si{M\odot}$ and the Eddington ratio is $\log \epsilon = -0.96$ \citep{Berton15a}, well within the typical black hole mass and Eddington ratio ranges for NLS1s.

It is one of the targets of a VLA survey of NLS1s carried out by \citet{Berton18} at $5\,\si{GHz}$.
In the literature, the object has been classified as radio-loud\footnote{The radio loudness of an AGN is estimated via the $R$ parameter, defined as the ratio between the radio $5\,\si{GHz}$ flux and the optical B-band flux \citep{Kellermann89}. For $R>10$ a source is considered radio-loud, otherwise it is considered radio-quiet.} \citep{Berton15a}.
An alternative classification, as radio-quiet, was instead reported by \citet{Doi13}.
This can be explained by the fact that Mrk 783 lies close to the threshold commonly used to distinguish between these two classes.
The calculation of the R parameter can be strongly influenced by the variability of the source, and by errors in measuring and defining the needed quantities \citep{Ho01}.

A KSR structure with a projected extension of the order of $10\,\si{kpc}$ was found in the preliminary analysis of the survey data by \citetalias{Congiu17}.
In the full resolution image, the emission appears as a diffuse structure located on the south-east side of the AGN nucleus, while in the tapered image it seems to be elongated also toward the opposite side \citepalias[Fig. 1,][]{Congiu17}.
\citet{Richards15} suggested that such extended emissions are not uncommon in radio-loud NLS1, but they seem to be less common in radio-quiet and moderately radio-loud ones. 
Only $28\%$ of the sample in \cite{Berton18} is classified as extended, and only a handful of these objects show extended emission which is comparable, in size and flux, to that found in Mrk 783.
The KSR in this galaxy is also characterised by a steep in-band spectral index ($\sim 2$, in the C-band).
The authors excluded the alternative of star formation, as a primary source of the radio emission \citepalias[Fig. 1,][]{Congiu17}, and tentatively classified the diffuse radio emission in this object as a relic due to the intermittent activity of the relativistic jet.

A preliminary optical follow-up of the object \citep{Congiu17c} revealed the presence of an extended narrow-line region (ENLR) possibly aligned with the radio emission and with a maximum extension of $\sim 30\,\si{kpc}$, one of the most extended ENLR discovered so far in the nearby Universe.
To our knowledge, this makes Mrk 783 the first NLS1 hosting both a KSR structure and an ENLR.

\section{Observations and Data Reduction}
\label{sec:data}

We summarise here the main details of the new e-MERLIN, VLBA and optical observations of Mrk 783, together with the data reduction procedures adopted. 

\subsection{e-MERLIN}
\label{sec:e-MERLIN}

We acquired new L band data with the e-MERLIN telescopes, to investigate the morphology of the extended radio emission on scales of $\sim0.2\,\si{arcsec}$ (program ID: CY6205, P.I.: Congiu).

The source was observed in two separate runs, January 6th, 2018, and January 16th, 2018, for a total on-source time of $13.2\,\si{hr}$.
These observations have a central frequency of $1.51\,\si{GHz}$, and a bandwidth of $512\,\si{MHz}$ divided in 8 spectral windows and $512$ channels per spectral window, which is the standard setup for e-MERLIN L-band observations.
During the data reduction, the channels have been averaged to 128 per spectral window.
A phase calibrator, a flux calibrator and a band-pass calibrator have been observed together with the main target during each observing run.

The e-MERLIN staff performed the calibration and preliminary imaging.
The e-MERLIN CASA pipeline v0.9 was used for the calibration of the data, while the self-calibration and the final imaging was performed with WSClean \citep{Offringa14,Offringa17}.
Since we expected the presence of extended structure in Mrk 783 and that the target was relatively faint, the self-calibration was not performed on the target itself but on another source in the field of view (FOV) of the telescope, and the final solutions applied to Mrk 783.

A Briggs weighting with a ROBUST parameter of 0.5 and a Gaussian taper with a full width at half maximum (FWHM) of $0.2\,\si{arcsec}$ have been used to extract a high-resolution image, while a Briggs weighting with a ROBUST parameter 2.0 and a $0.5\,\si{arcsec}$ Gaussian taper were used to extract an image with a spatial resolution similar to that of the VLA image described in \citetalias{Congiu17}.
Each image has also been cleaned using the \verb!-joinchannel! option in WSClean, which allows one to obtain simultaneously the total image and the images in each of the four bands observed by the instruments.
We used these images to estimate the in-band spectral index of the radio emission (Sec.\,\ref{sec:results}).

\subsection{VLBA}
\label{sec:vlba}

We observed Mrk 783 with the VLBA at L and C bands, to investigate the properties of the radio emission at high spatial resolution (Program ID: BC247, P.I.: Congiu).

The observations were performed on May 4th, 2018 (C band) and May 21st, 2018 (L band) for a total on-source time per band of $36\,\si{min}$.
The central frequencies for these observations are $1.57\,\si{GHz}$ and $4.98\,\si{GHz}$ for the L-band and C-band respectively, the bandwidth is $128\,\si{MHz}$ and it is divided in 4 channels.
The target was observed in phase-referencing mode with a 3:2 target/phase calibrator cycle.
3C\,345 (R.A. = $16$h $42$m $58.8$s Dec=$+39$d $48$m $37.0$s) was observed as a fringe finder for the experiment, while the VLBA calibrators J1300+141A (R.A. = $13$h $00$m $20.9$s Dec=$+14$d $17$m $18.5$s) and J1300+141B (R.A. = $13$h $00$m $41.0$s Dec=$+14$d $17$m $29.41$s) were observed, respectively, as a phase-reference and a phase-check calibrator.

The data reduction was performed with the AIPS standard VLBA pipeline (VLBARUN).
Since the target was faint in both the observed bands, we did not perform any self-calibration.
The final cleaning and imaging were performed with CASA, using a natural weighting scheme for both frequencies which yielded a beam size of $14\times6.7\,\si{mas^2}$ and $3.9\times1.6\,\si{mas^2}$ for the L-band and C-band respectively.

\subsection{Optical Images}

We acquired V band images of the galaxy with the du Pont telescope of the Las Campanas Observatory on April 7th, 2019.
The data were reduced using standard IRAF procedures. 
The raw images have been corrected for overscan and flat-field. 
They have been astrometrized using Astrometry.net\footnote{\url{http://astrometry.net/}} \citep{Lang10}. 
No flux calibration was applied. 
The final image is the combination of 3 frames, for a total exposure time of $900\,\si{s}$ on-source, and an average seeing of $0.86\,\si{arcsec}$.

\section{Results}
\label{sec:results}
\begin{figure*}
    \centering
    \includegraphics[width=0.9\textwidth]{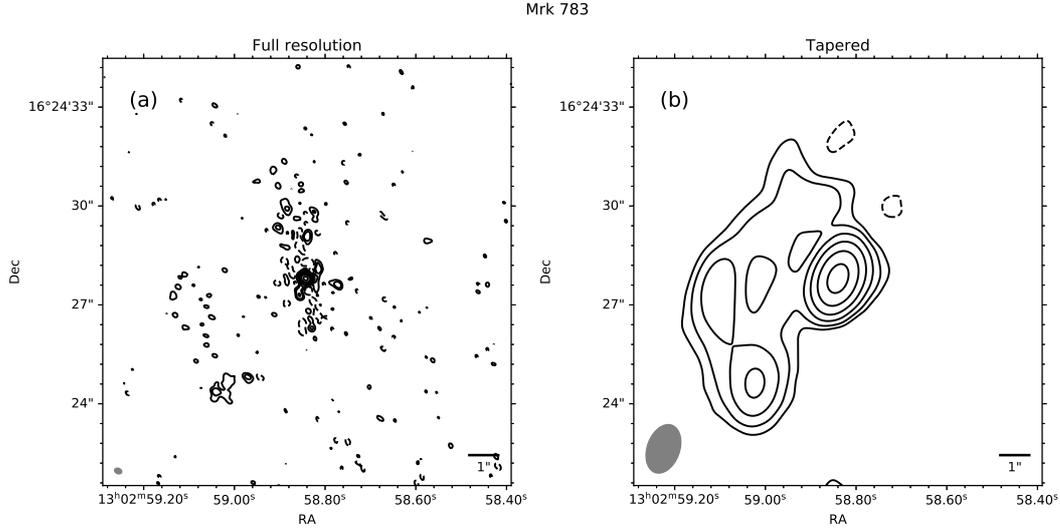}
    \caption{e-MERLIN L band images of Mrk 783. \textbf{Panel a} shows the high resolution image (Briggs weighting, robust parameter 0.5, Gaussian taper with a FWHM of $0.2\,\si{arcsec}$), while \textbf{panel b} shows the low resolution image (Briggs weighting, robust parameter 2.0, Gaussian taper with FWHM $0.5\,\si{arcsec}$).
    In both images the contours are at [-3, 3, 6, 12, 24, 48, 92] $\times\sigma$, where $\sigma$ is the rms of the image measured in regions where no significant emission is visible. $\sigma$'s value is  $0.025\,\si{mJy.beam^{-1}}$ and $0.061\,\si{mJy.beam^{-1}}$ for the full resolution and tapered image respectively. Dashed contours represent negative values.
    Finally the beam size is $0.24 \times0.17\,\si{arcsec^2}$ for the image on the left and $1.53 \times1.02\,\si{arcsec^2}$ for the image on the right. In both images the beam is shown on the bottom left corner.}
    \label{fig:e-MERLIN_images}
\end{figure*}

\begin{figure*}
    \centering
    \includegraphics[width=0.9\textwidth]{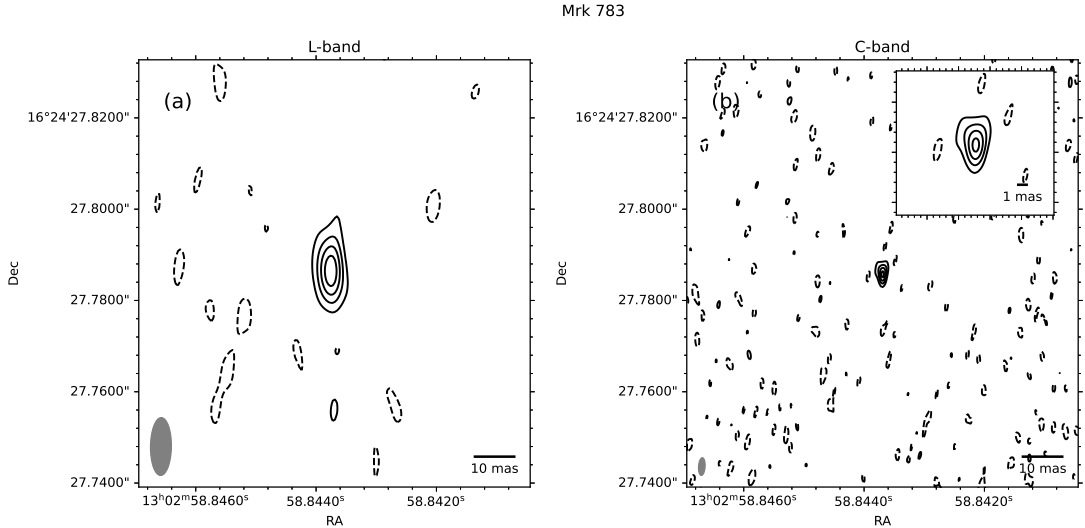}
    \caption{VLBA L-band (\textbf{panel a}) and C-band (\textbf{panel b}) images of Mrk 783. The images are on the same scale. The inset in the upper-right corner of the right panel shows a zoom on the C-band emission. The contours of both images are at [-2, 3, 6, 9, 12] $\times\sigma$ where sigma is defined as in Fig.\,\ref{fig:e-MERLIN_images} and it is $0.047\,\si{mJy.beam^{-1}}$ and $0.022\,\si{mJy.beam^{-1}}$ for the L- and C-band images respectively. Dashed contours represent negative values. The beam size is $14\times6.7\,\si{mas^2}$ for the image on the left and $3.9\times1.6\,\si{mas^2}$ for the image on the right. In both images the beam is shown on the bottom left corner.}
    \label{fig:VLBA_images}
\end{figure*}

In this section we describe the main results and measurements obtained by analysing our new radio observations of Mrk 783.

\subsection{Morphology}
\label{sec:morphology}
Fig.\,\ref{fig:e-MERLIN_images} and Fig.\,\ref{fig:VLBA_images} show the new e-MERLIN and VLBA images of Mrk 783, respectively.
The source is detected with both instruments and in all the observed bands.
The full resolution e-MERLIN image (Fig.\,\ref{fig:e-MERLIN_images}, panel a) clearly shows the presence of a core, while, due to the absence of short baselines, the majority of the extended emission observed in \citetalias{Congiu17} is resolved out.
The minimum baseline of the array is $11\,\si{km}$, which means that all the structures larger than $\sim 5\,\si{arcsec}$ ($\sim6.5\,\si{kpc}$ at z $=0.0672$) are resolved out by the instrument.
Therefore, only traces of the diffuse emission can be observed, but their distribution seems to follow what we observed in the VLA image.
The morphology of the radio emission observed in the e-MERLIN tapered image (Fig.\,\ref{fig:e-MERLIN_images}, panel b) is, indeed, remarkably similar to the top right panel of Fig.\,1 in \citetalias{Congiu17}, with a relatively bright core and a diffuse emission extending towards the south-east side of the nucleus.

The VLBA L-band image shows an unresolved core, similar to what was previously observed by \citet{Doi13}.
On the other hand, the C-band image shows a partially resolved structure located at position angle (PA) of $\ang{60}$.
This structure has not been previously observed, and its direction is $\sim\ang{70}$ north with respect to the axis of the large scale structure observed in the e-MERLIN images and the VLA images from \citetalias{Congiu17}.

\subsection{Flux Densities}
\label{sec:fluxes}

The total flux density has been measured, in all images, summing the flux density of the pixels inside the $3\sigma$ contour, where $\sigma$ is the rms of the image measured in a region far away from the source.
The error on the total flux density has been estimated as the rms times the square root of the area covered by the $3\sigma$ contour expressed in units of beams \citep[see, e.g., Sec. 3 in][]{Panessa15}.
The core properties are recovered using the CASA \verb!imfit! routine, which produces both measurements and errors.
Tab.\,\ref{tab:measurement} summarises the properties of the images and the measured quantities.
We used a single component to fit the core in all images except for the VLBA C-band image (Fig.\,\ref{fig:VLBA_images} panel b), where we modelled the emission with 2 Gaussian components.
We limited the fit only to the region corresponding to the $3\sigma$ contour, and we manually estimated the first guess parameters required by the task to fit each component (peak intensity, position, size and PA).
To test the significance of the fit, we also tried to reproduce the same emission using a single Gaussian component.
The results of the multi-Gaussian fit and of the single Gaussian fit of the C-band image are shown in Tab.\,\ref{tab:mgauss} and in Fig.\,\ref{fig:fit}.
A single Gaussian is still able to reproduce the emission, but the errors on the position and the flux density are significantly larger with respect to the two Gaussians fit.
Also, the rms of the residual image when using a single component is higher ($17\,\si{\micro Jy/beam}$) with respect to that of the two Gaussians fit ($6\,\si{\micro Jy/beam}$).
Finally, it is possible to see in Fig.\,\ref{fig:fit} how the single Gaussian fit reproduced the observations poorly compared to the 2 Gaussians fit.
In particular, after the subtraction of the fitted model from the data, a $3\sigma$ residual is left at the position of the resolved component, while the core is evidently over subtracted, as shown by the negative contours in the lower left panel of Fig.\,\ref{fig:fit}.

\begin{figure*}
    \centering
    \includegraphics[width=\textwidth]{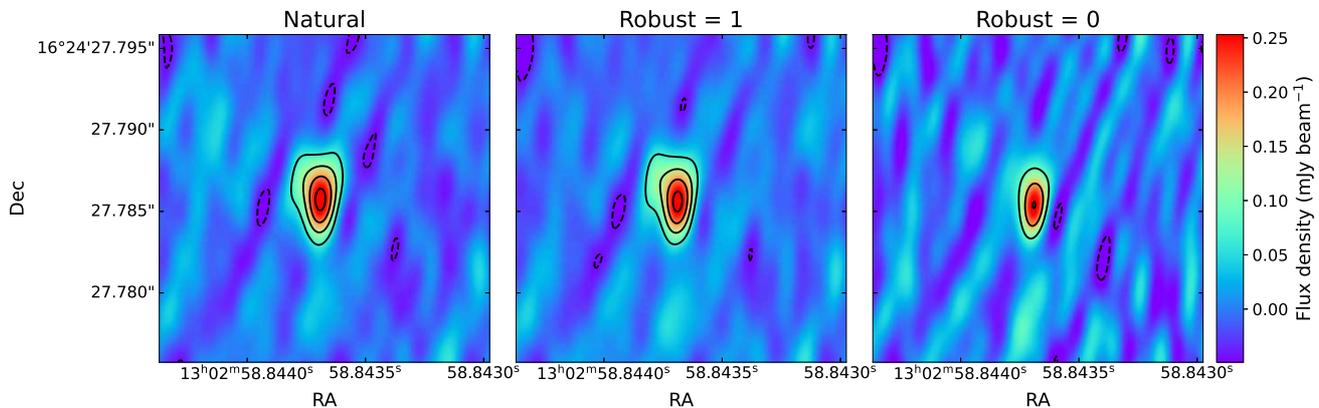}
    \caption{C-band images of Mrk 783 recovered from the VLBA data using different weighting schemes. The image on the left is the same image shown in the left panel of Fig.\,\ref{fig:VLBA_images}. The image on the centre is obtained using a \emph{briggs} weighting scheme with robust $=1$ , the image on the right uses robust $=0$. The contours levels are the same as in the left panel of Fig.\,\ref{fig:VLBA_images}, but for the left and centre panels $\sigma=2.2e-5\,\si{mJy.beam^{-1}}$, while for the right panel $\sigma=3.0e-5\,\si{mJy.beam^{-1}}$. These values of $\sigma$ were measured in regions of the images free from emission.} 
    \label{fig:multi_weight}
\end{figure*}

\smallskip
Even though the 2 Gaussians fit better describes the data, the elongation of the source, here interpreted as a pc-scale jet, is extremely faint and only barely resolved, therefore in the following we discuss the possibility that we are dealing with an image artifact.
We tried to recover several images using different weighting schemes, to see if it was possible to detect the second component also with weighting schemes different from the natural one.
In Fig.\,\ref{fig:multi_weight} we show some of these images. 
The first panel on the left shows the same image as in the panel b of Fig.\,\ref{fig:VLBA_images}, which is our reference image, and it was recovered with a natural weighting scheme.
The image on the centre panel uses instead a Briggs weighting scheme with robust $=1$. 
This combination returns an image which is very similar to the natural one, and the second component is clearly detected.
This is expected because with such a robust parameter this weighting scheme is close to a natural weighting scheme. 
The right panel shows the image recovered using a Briggs weighting scheme with robust $=0$.
This image is much more noisy, and the object is not as resolved as in the other two images.
However, the central component is not perfectly symmetric and a hint of extension can be identified at the position angle where we expect to see the second component.
Such a behaviour is expected because with robust $=0$, this weighting scheme is more focused in producing images with a small beam size, but at the expense of a decreased sensitivity, as proven by the increased rms of the image.

Considering the new images obtained with different weighting, the multi-Gaussian fit presented at the beginning of this section and in Fig.\,\ref{fig:fit}, and the measured properties discussed in the following paragraphs, we are confident that this feature is not an artefact due to the data reduction, but it is, indeed, a real structure.
Therefore, the radio emission of Mrk 783 at VLBA scales is composed of a core and a second component, seemingly a pc-scale jet.
New observations, with higher sensitivity are, however, required to better characterise the emission and definitively confirm the nature of the second component.

\subsection{Spectral indexes and brightness temperatures}
\label{sec:spind}

Since we detected Mrk 783 both in the C-band and L-band, we can estimate the spectral index of the core emission.
Unfortunately, the observations were not simultaneous, and this can introduce biases when measuring the spectral index.
The steep core spectral index measured by \citetalias{Congiu17} already excluded the possibility of Mrk 783 hosting a relativistic jet beamed toward the Earth.
Highly beamed sources are usually characterised by flat or inverted spectral index \citep{Fan10}, and by extremely high brightness temperatures \citep[$>10^{10}$ K, ][]{Kovalev05}, which are not observed in Mrk 783 neither by us (see Sec.\,\ref{sec:discussion1}), nor by previous works \citep[see][]{Doi13}.
As a consequence, even if long scale (years) variability can be expected when observing the nucleus, the source should be relatively stable on timescales of days or few weeks.
Our observations were only 17 days apart. 
Therefore, we can assume that the spectral index measured from the VLBA data is a reasonable estimate.
Considering the total flux density recovered by the fits in both bands\footnote{For the C-band we considered the fitted flux densities of both components.}, we obtained $\alpha_{VLBA} = 0.63\pm0.09$.

The difference in the scales observed by the VLA and the e-MERLIN array, which results in a consistent loss of flux in the e-MERLIN image (Sec.\,\ref{sec:discussion1}), does not allow us to measure a spectral index of the KSR taking advantage of the large difference in frequencies.
On the other hand, estimates of the in-band spectral indexes of the extended emission using the e-MERLIN data, similar to those from \citetalias{Congiu17}, are unreliable, both because of the small width of the observed band (only $512\,\si{MHz}$) and because most of the extended emission is resolved out.
This translates to an inaccurate measurement of the flux from the tapered, low resolution, images.
This second issue, however, should not affect the estimation of the core spectral index, which can then be estimated from the images of the four observed sub-bands.
To do so, we fitted each one of the four full resolution images with a Gaussian to measure the core flux density, and we obtained an in-band spectral index $\alpha_c = 0.89\pm0.15$.
We stress, however, that this estimate is still affected by the small bandwidth, so it should interpreted as a trend, more than as an absolute value.

Finally, using the deconvolved sizes and flux densities recovered by the CASA fitting routine from the VLBA images, we also measured the brightness temperature of Mrk 783 core in the L-band image and of both the core and the extended component in the C-band one (Tab.\,\ref{tab:tb}).
Both components in the C-band are not resolved, and we used the convolved size of the source to estimate their brightness temperature, which, therefore, is only a lower limit.

\begin{table*}
    \centering
    \caption{Principal properties of the new images of Mrk 783. The columns are: (1) name of the image, (2) reference frequency, (3) the bandwidth of the observations, (4) size of the beam, (5) rms of the image measured in an empty area far from the object, (6) maximum flux density per beam measured in the image, (7) flux density contained in the $3\sigma$ contour, (8) flux density of the core as measured by the CASA viewer fitting algorithm.}
    \label{tab:measurement}
    \begin{tabular}{cccccccc}
    \hline
    Image&Frequency&Bandwidth&Beam Size & $\sigma$            &Peak            &Total Flux& Core Flux\\
         & GHz  &MHz &  & $\si{mJy/beam}$& $\si{mJy/beam}$&$\si{mJy}$&$\si{mJy}$\\
    (1) & (2) & (3) & (4) & (5) &(6)&(7)&(8)\\
    \hline
    e-MERLIN high-res& 1.51&512&$0.24\times0.17\,\si{arcsec^2}$& $0.025$& $5.924$&$12.3\pm1.2$ &$7.293\pm0.053$\\
    e-MERLIN low-res & 1.51&512&$1.53\times1.02\,\si{arcsec^2}$& $0.061$& $7.188$&$14.9\pm1.1$ &$7.49\pm0.24$  \\
    VLBA L-band     & 1.57&128 & $14\times6.7\,\si{mas^2}$      & $0.047$& $0.681$&$0.614\pm0.086$ &$0.764\pm0.035$\\
    VLBA C-band     & 4.98&128 &$3.9\times1.6\,\si{mas^2}$     & $0.022$& $0.287$&$0.278\pm0.042$ &-\\
    \hline
    \end{tabular}

\end{table*}

\begin{table*}
\centering
    \caption{Results of the multi-Gaussian fit of the VLBA C-band image. The columns are: (1) fitted component, (2) right ascension (3) declination, (4) pixel size, (5) separation between the two components in milliarcseconds and (6) in parsecs, considering a redshift of $0.0672$, (7) deconvolved major axis, (8) deconvolved minor axis, (9) peak flux density, (10) integrated flux density.}
    \label{tab:mgauss}
    \begin{tabular}{lccccccccc}
    \hline
    Component& RA & Dec& Pixel &\multicolumn{2}{c}{Separation}& Major Axis & Minor Axis& Peak &Flux Density\\
         &   pix & pix & mas&mas&pc& mas&mas &$\si{\micro Jy/beam}$& $\si{\micro Jy}$\\
    \multicolumn{1}{c}{(1)} & (2) & (3) & (4) & (5)& (6) & (7)&(8)&(9)&(10)\\
    \hline 
    core$^a$ & 559.22$\pm$0.20 & 510.29$\pm$0.93 &0.1 &-&-&-&-& 171.8$\pm$4.8& 175$\pm$10\\
    extended$^b$  & 551.10$\pm$0.27 & 515.02$\pm$0.79 &0.1 &0.9&1.2& - & - &158.3$\pm$5.8 &193$\pm$12\\
    core$^c$ & 555.98$\pm$0.42 & 511.24$\pm$1.45 &0.1 & - & - &1.58$\pm$0.68&0.85$\pm$0.65& 269$\pm$17& 374$\pm$39\\
    \hline
    \multicolumn{9}{l}{$^a$The source is a point source.}\\
    \multicolumn{9}{l}{$^b$The source might be resolved only in one direction.}\\
    \multicolumn{9}{l}{$^c$Result of the fit performed with a single Gaussian.}\\
    \end{tabular}
\end{table*}

\begin{table*}
\centering
    \caption{Core brightness temperature measured from the VLBA images.
    To calculate the brightness temperature, we used the deconvolved size for the L-band component, and the convolved size for the two components of the C-band image.
    The columns are: (1) band, (2) analysed component, (3) reference frequency, (4) flux density recovered from the fit, (5) major axis, (6) minor axis, (7) brightness temperature.}
    \label{tab:tb}
    \begin{tabular}{ccccccc}
    \hline
    Band&Component &Frequency& Flux Density& Major Axis& Minor Axis& T$_B$\\
         & &   GHz   & $\si{mJy}$& $\si{mas}$&$\si{mas}$&$\si{K}$\\
    (1) & (2) & (3) & (4) & (5) &(6)&(7)\\
    \hline
    L     &-& $1.57$ &$0.764$ & $3.5\pm1.4$& $1.2\pm1.1$&$1.55\times10^8$\\
    C$^a$ &core&$4.98$ &$0.175$ & $4.47\pm0.22$ &$1.373\pm0.021$ &$\geq2.2\times10^6$\\
    C$^a$ &extended&$4.98$ &$0.193$ & $3.80\pm0.19$&$1.937\pm0.053$&$\geq 1.9\times10^6$\\
    \hline
    \multicolumn{6}{l}{$^a$The source is not resolved, hence the brightness temperature is a lower limit.}
    \end{tabular}
\end{table*}

\begin{figure*}
    \centering
    \includegraphics[width=\textwidth]{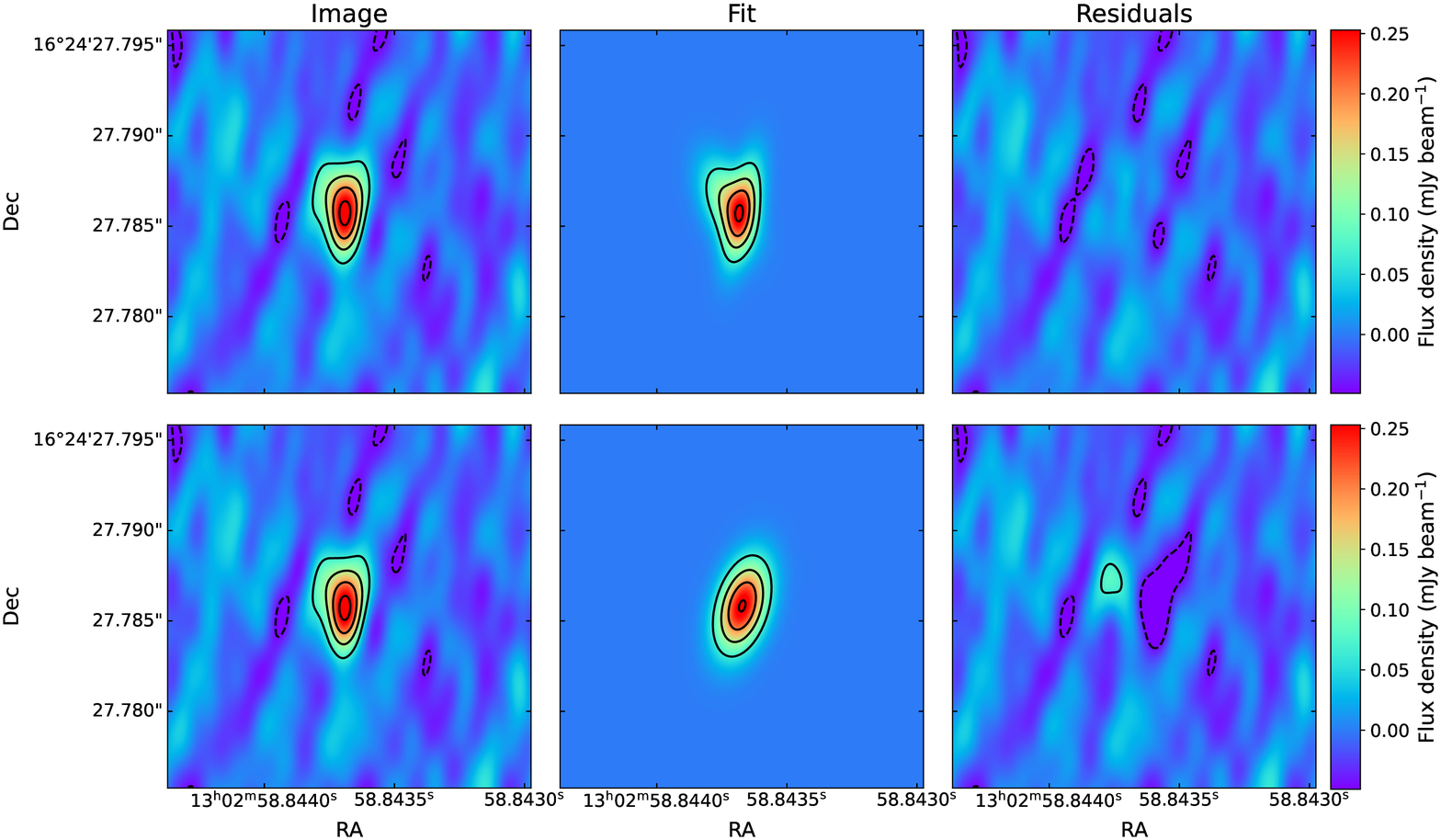}
    \caption{Results of the fits of the VLBA C-band image. In each row there are, from left to right: the original image, the fitted model and the residual from the subtraction of the model from the original image. The contour levels are the same as in Fig.\,\ref{fig:VLBA_images}, panel b. The first row shows the results of the fit with 2 Gaussian components, representing the core and the putative jet. The second row shows the results of the single component fit.}
    \label{fig:fit}
\end{figure*}

\section{Discussion}
\label{sec:discussion}

\subsection{Radio properties}
\label{sec:discussion1}

The new e-MERLIN images confirm the results of \citetalias{Congiu17}.
The galaxy appears as a compact core with a diffuse emission on the south-east side also at the higher resolution.
There is no compact feature significantly detected in the high-resolution e-MERLIN image (Fig.\,\ref{fig:e-MERLIN_images}, panel a) in the region corresponding to the VLA extended emission.
As a consequence, the intrinsically diffuse nature of such emission is strongly favoured.
The in-band spectral index of the core obtained from the e-MERLIN data is steep ($\alpha_c = 0.89\pm0.15$) and consistent, within the uncertainties, with that found by \citetalias{Congiu17} ($\alpha = 0.67\pm0.13$) and slightly steeper than the spectral index obtained from the VLBA images ($\alpha_{VLBA} = 0.63\pm0.09$).
An estimate of the in-band spectral index for the extended emission or spectral indexes derived from the new e-MERLIN data and the VLA data from \citetalias{Congiu17} would not be reliable for the reasons explained in Sec.\,\ref{sec:spind}.

We also used our VLBA images in the L-band and in the C-band, to investigate the properties of the target at sub-arcseconds scales.
Mrk 783 has already been observed in L-band with the VLBA by \citet{Doi13} in 2005.
They observed an unresolved component, with a flux density of $1.3\pm0.2\,\si{mJy}$ at a reference frequency of $1.7\,\si{GHz}$.

Our new VLBA L-band image is very similar to Doi's, only showing an unresolved component.
We can compare our flux density measurement in this band to the one in \citet{Doi13} to see if there has been some changes. 
However, the reference frequency of Doi's observation was slightly different and to do a direct comparison we have to first correct for this discrepancy.
Our flux density at $1.5\,\si{GHz}$ is $0.764\pm0.035\,\si{mJy}$ and the measured spectral index is $\alpha = 0.63$ (Sec.\,\ref{sec:spind}).
Considering that $S_{\nu1}/S_{\nu2} = (\nu1/\nu2)^{-\alpha}$, the expected flux density at $1.7\,\si{GHz}$ is, therefore, $0.73\pm0.07\,\si{mJy}$.
This is only $\sim 56\%$ of what measured in \citet{Doi13}.
Radio-loud NLS1s \citep{Gu15} and several other classes of radio-loud AGN \citep[e.g.][]{Hovatta08} are known to be variable sources, especially on small spatial scales such as those observed by the VLBA.
However, strong variability on timescales of several years has also been observed in the nuclei of radio-quiet sources \citep[][and references therein]{Mundell09}.
Mrk 783 is a borderline object, at the edge between radio-quiet and radio-loud ones, hence expecting a certain degree of variability on long timescales is everything but farfetched.

Interestingly, this is the first time that this source has been observed at VLBA resolution in the C-band.
The image shows the presence of a single, partially resolved, component which is elongated at PA $\sim\ang{60}$.
The fits in Sec.\,\ref{sec:fluxes}, confirm the presence of a second component close to the radio core.
We proposed that this might be a small-scale jet produced by the AGN.
The beam size of the L-band image is almost three times larger than that of the C-band image, which is the reason why the structure is not resolved in our L-band image.
Also, the beam size of the old VLBA image from \citet{Doi13} is twice that of our C-band image, which is still not enough to resolve such a tiny extended structure.

In Sec.\,\ref{sec:spind} we measured the spectral index using the flux density of the source in our VLBA L-band and C-band images ($\alpha_{VLBA} = 0.63\pm0.09$).
This represents the first estimate of the spectral index at these scales (or frequencies), therefore we have no comparison with previous similar measurements.
However, this value is consistent with the core in-band spectral index measured from the VLA image in \citetalias{Congiu17} ($\alpha = 0.67\pm0.13$).
Steep radio cores at VLBA scales are not uncommon in Seyfert galaxies \citep{Orienti10,Kharb10}, and they usually indicate the presence of optically thin synchrotron emission.

We estimated the brightness temperature of the radio emission in both VLBA bands using Eq.\,\ref{eq:tb}  \citep{Doi13}:
\begin{equation}
\label{eq:tb}
T_B=1.8\times 10^9 (1+z)\frac{S_{\nu,c}}{\nu^2\phi_{maj}\phi_{min}}\sim7600 {\rm\, K},
\end{equation}
where $z$ is the redshift, $\phi_{maj}$ is the object major axis, and $\phi_{min}$ is the minor axis.
The measurements are reported in Tab.\,\ref{tab:tb}.
For the C-band, we used the results of the multi-Gaussian fit (Tab.\,\ref{tab:mgauss}) to estimate the brightness temperature of both the core and the extended component separately.
As the two components are not resolved, so we were able to estimate only their convolved size which represents an upper limit to their real dimension.
For this reason, the measured brightness temperature is only a lower limit.
In all cases, the brightness temperature ($1.55\times10^8\,\si{K}$ in the L-band, $>10^6\,\si{K}$ for both components in the C-band) is consistent with the values found by \citet{Nagar05} for other AGN.
Therefore, it is highly probable that the observed structure is a small-scale jet launched by the AGN.
A small scale active jet in its nucleus, could also contribute to the observed variability of the source.

\subsection{Precession or intermittent activity?}
\label{sec:discussion2}

In the previous section, we reported the discovery of a pc-scale extended radio emission in the VLBA C-band image. From the physical properties we were able to estimate, we suggested that this extended emission is a pc-scale jet produced by the AGN.

There are two interesting properties of this putative pc-scale jet: it has been observed in a galaxy with an extended emission that has been previously classified as a relic, and the small-scale jet and the large scale radio emission are not aligned.
The small scale jet is located at a PA $\sim \ang{70}$ north with respect to the axis of the KSR structure.

In the following, we want to discuss two possible scenarios that can result in the production kpc-scale relic(-like) emission and a misaligned pc-scale jet.
The two scenarios are: i) precession and ii) intermittent activity.
The former scenario is a continuous precession of the jet axis, i.e. the rotation of the jet axis as a function of time.
Such phenomenon produces extended radio structures with peculiar morphologies, bent or S-shaped, and they are commonly observed in Seyfert galaxies \citep{Kukula95,Nagar99,Thean00,Kharb06,Kharb10}.
The presence of a binary SMBH is one of the most probable causes for jet precession \citep[e.g.][]{Roos88}, but other phenomena such as the presence of a warped disk \citep{Pringle96,Pringle97}, the precession of an elliptical disk \citep{Eracleous95} and many others have been suggested in the literature.
The morphology of the extended radio emission in Mrk 783 can resemble an S, especially considering the tapered image in \citetalias{Congiu17}. 
Therefore, precession is a possibility that cannot be excluded.

\begin{figure}
    \centering
    \includegraphics[width=0.45\textwidth, ]{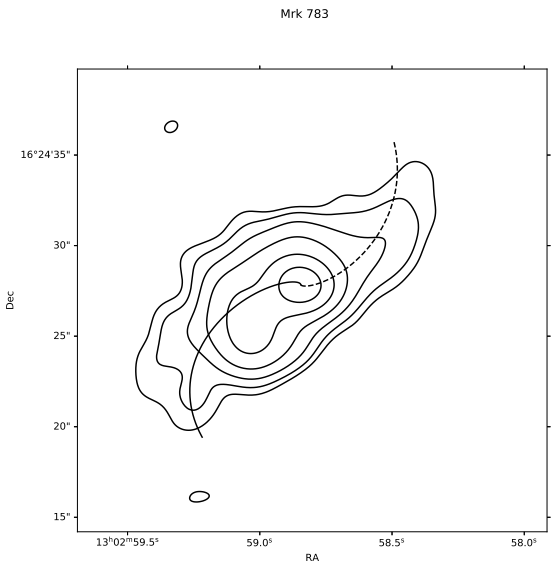}
    \caption{Tapered VLA C-band image of Mrk 783 from \citetalias{Congiu17} with the results from the model by \citet{Hjellming81}. The solid curve represent the jet, while the dashed one the counter-jet.}
    \label{fig:precession}
\end{figure}

\citet{Hjellming81} developed a precession model to study jets produced by galactic X-ray binaries, which has been successfully used to fit AGN jets \citep[e.g.][]{Kharb06,Kharb10}.
In Fig.\,\ref{fig:precession} we show the result of the best model reproducing the radio emission and with jet inclination and velocity compatible with the constraint previously obtained. 
The model predicts a clockwise\footnote{The precessing direction is defined for an observer located in the galaxy nucleus and looking in the direction of movement of the jet particles} preceding jet, with a jet velocity $\beta_m = 0.27$, a jet precession axis inclination $\theta_m = \ang{45}$, a precession period of $5\times10^5\,\si{yrs}$, a precession cone half-opening angle $\phi = \ang{54}$ and a PA $=\ang{192}$.

In the second scenario, the jet is reactivated after a period of inactivity.
In this case, the KSR structure is the relic of a previous activity period, as suggested in \citetalias{Congiu17}, while the small-scale jet started its activity only recently and at a different position angle.
This kind of situation has been invoked to explain the morphology of different kinds of radio sources \citep[e.g.][]{Schoenmakers00a,Kaiser00,Kharb06,Murgia11,Orru15}.
While most of the literature agrees that the interruption of the jet activity is due to the variation or interruption of the activity of the central AGN, the physical processes producing this change in the activity status are still debated.
The two most widely accepted models are the accretion of new gas due to galaxy mergers \citep{Lara99,Schoenmakers00a}, and chaotic accretion \citep{Hernandez17}.
The infall of large masses of gas into the vicinity of the SMBH could cause at first an instability of the accretion flow, reducing or stopping the AGN and jet activity \citep{Schoenmakers00a}.
However, once the gas is redistributed and the flow is regular, the AGN and the jet start over again.
Since there is not a preferential direction for the infall of new gas, the orientation of the new jet can be different from the original one, producing structures like the ones in Mrk 6 \citep{Kharb06}.
Another possibility for the interruption of the jet activity is an internal instability of the accretion flow, e.g. due to warping of the accretion disk \citep{Pringle96,Pringle97}.
This kind of instability can change the orientation of the jet and, if it is not strong enough to terminate the jet activity, it can also produce precession, as we mentioned in the previous paragraph.

\begin{figure}
    \centering
    \includegraphics[width=0.48\textwidth]{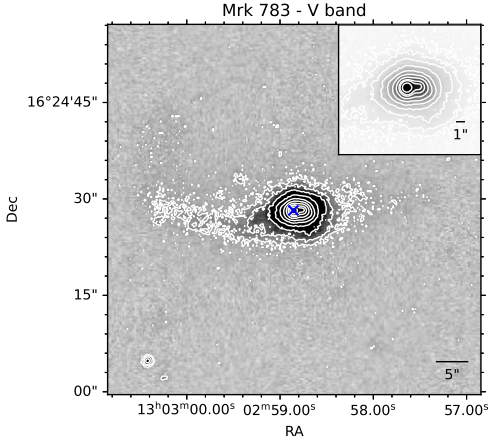}
    \caption{V band image of Mrk 783 acquired with the du Pont telescope of the Las Campanas Observatory. The blue cross shows the optical position of the AGN. The white contours represent the $[3, 6, 12, 24, 48, 92, 184, 368, 552, 1104]\times \sigma_{sky}$, where $\sigma_{sky}$ is the standard deviation of the sky (in counts) measured in a region without significant emission. The plot on the upper right corner of this panel shows a zoom in on the galaxy nucleus with a different colour scale, to highlight the shape of the isophote.}
    \label{fig:lco}
\end{figure}

We acquired a V-band image of the galaxy with the du Pont telescope at the Las Campanas Observatory to study the morphology of the object and look for signs of interaction with other galaxies. 
Fig.\,\ref{fig:lco} shows the V-band optical image of Mrk 783 (both grey-scale and contours).
The galaxy shows a disturbed morphology.
A low surface brightness extended structure can be observed on the left side of the nucleus (Fig.\,\ref{fig:lco}).
A similar, but fainter and less extended structure can also be observed on the right side of the nucleus.
While this emission can be produced by very loose spiral arms, their morphology resembles more that of tidal tails observed, although often more prominently in interacting galaxies, i.e. Antennae galaxies.

The contours of the central region of the galaxy in Fig.\,\ref{fig:lco} have a complex shape.
They show the presence of the AGN, which is marked with a blue cross, and of a second, fainter, point-like component whose origin is still not clear.
The distance between the two structures is $1.2\pm0.1\si{arcsec}$, corresponding to $\sim1.5\,\si{kpc}$.
No radio emission is observed at this position in the e-MERLIN images, while the VLA images from \citetalias{Congiu17} show some diffuse emission on the west side of the nucleus that might be connected to this structure.
In the hypothesis of a recent merger, this secondary component might be the nucleus of the companion galaxy, but a more detailed study is needed to confirm its nature.

In conclusion, both scenarios presented in this section seem to be suitable to explain our radio observations of Mrk 783.
The jet precession model from \citet{Hjellming81} is able to reproduce the morphology of the large scale emission observed in \citetalias{Congiu17}, while the optical morphology of the galaxy suggests that the galaxy underwent a recent merger.
The disturbances caused by the merging event can result in the inflow of new gas in the AGN area.
The inflow of this new gas can introduce instabilities in the region close to the SMBH, and, depending on their characteristics, they can result in the interruption of the AGN activity or in the precession of the jet.

\section{Conclusions}
\label{sec:conclusions}

In this paper, we have analysed new multi-band e-MERLIN and VLBA observation of the NLS1 galaxy Mrk 783.
The galaxy is known to host a KSR structure which has been previously classified as a relic possibly caused by an intermittent activity of the AGN \citepalias{Congiu17}.
The tapered L-band e-MERLIN image is very similar to the VLA image presented in \citetalias{Congiu17} while the high-resolution image confirms the diffuse nature of the structure.
From these data, we measured the in-band spectral index, which is consistent with what \citetalias{Congiu17} measured in the VLA C-band image.

The object is detected in both the VLBA bands observed.
The VLBA L-band image shows an unresolved core, similar to what was found by \citet{Doi13}, but we measure only 50\% of the flux density reported in Doi's paper, a possible consequence of a strong variability on timescales of a decade or more.
In addition, the C-band image shows a partially resolved structure at PA$ = \ang{60}$, about $\ang{70}$ north with respect to the axis of the KSR structure.
From these new data, we measured a spectral index of $0.63\pm0.09$, which is consistent with the value measured by \citetalias{Congiu17} for the core of the emission in the VLA image and with the presence of a small-scale, extended structure \citep{Orienti10,Kharb10}.

The brightness temperature of the core in both VLBA images is $>10^6\,\si{K}$, in the expected range for non thermal radio emission typical of AGN.
Also the brightness temperature of the extended structure in the C-band is of the same order of magnitude, suggesting that the emission is produced by an AGN jet.

We investigated two scenarios that can explain the presence of a KSR structure and a misaligned pc-scale jet: precession and intermittent activity.
We successfully fitted the tapered VLA image from \citetalias{Congiu17} with the \citet{Hjellming81} jet precession model, which confirms that jet precession is a plausible scenario.
We have also analysed a new optical image of the target, which reveals the possibility of a recent interaction with a companion object.
Interaction is one of the main causes of the intermittent activity of AGN and jets \citep{Schoenmakers00b}, but it can also create instabilities which may result in jet precession.
Low surface brightness extended structures, similar to tidal tails observed in interacting galaxies, are detected on both sides of the galaxy nucleus in the optical V-band image.
The isophotes of the internal part of the galaxy indicate the presence of a second point-like structure, close to the AGN, which might be the nucleus of the second galaxy involved in the proposed merger.
In conclusion, while precession and intermittent activity are two different processes they are often produced by similar causes and they can produce similar results and we have no decisive elements, so far, which can be used to determine which one of the two scenarios is the closest to reality.

New, multiwavelength radio observation can be used to produce a robust, spatially resolved map of the spectral index of the KSR structure.
Studying how the spectral index is changing spatially may help us to identify the true nature of the extended radio emission.
Higher resolution and deeper radio observations are recommended to provide a definite confirmation of the presence of the pc-scale jet, while a radio monitoring program is planned to study its evolution and the flux variability of the radio emission at milliarcsecond scale.
Finally, deep optical integral field observations of the galaxy can help clarify the nature of the two extended structures and the second point-like component close to the AGN, while allowing a detailed study of the ENLR reported in the literature by \citet{Congiu17c}.

\section*{Acknowledgement}
The authors of the paper want to thank the anonymous referee for helpful suggestions and  Dr. Javier Molton of the e-MERLIN support centre for his precious help on the data reduction.
e-MERLIN is a National Facility operated by the University of Manchester at Jodrell Bank Observatory on behalf of STFC.
E.C. acknowledges support from ANID project Basal AFB-170002.
The National Radio Astronomy Observatory is a facility of the National Science Foundation operated under cooperative agreement by Associated Universities, Inc.
This paper includes data gathered with the 2.5 meter du Pont Telescopes located at Las Campanas Observatory, Chile. 
This research has made use of the NASA/IPAC Extragalactic Database (NED), which is operated by the Jet Propulsion Laboratory, California Institute of Technology, under contract with the National Aeronautics and Space Administration.
This research made use of facilities at the Mets\"ahovi Radio Observatory, operated by the Aalto University, Finland.
This research made use of Astropy,\footnote{http://www.astropy.org} a community-developed core Python package for Astronomy \citep{Astropy13,Astropy18}. 
This research made use of APLpy, an open-source plotting package for Python hosted at \url{http://aplpy.github.com}.

\section*{Data availability}

The data underlying this article will be shared on reasonable request to the corresponding author.


\bibliographystyle{mnras}
\bibliography{bibliografia_def} 




\bsp	
\label{lastpage}
\end{document}